\documentclass[sigconf]{acmart}
\usepackage{amsmath}
\usepackage[utf8]{inputenc}
\usepackage{multirow}
\usepackage{subfig}
\usepackage{hyperref}
\usepackage{subcaption}
\usepackage[noabbrev,capitalise,nameinlink]{cleveref}

\definecolor{SubtleColor}{rgb}{0,0,.50}
\newcounter{margincounter}

\begin{document}
\title{Soft Curriculum Learning for Optimizing Fresh and Generalized Recommendations}

\author{Arnab Bhadury}
\authornote{Equal Contribution}
\affiliation{%
  \institution{Google}
  \city{Vancouver}
  \state{BC}
  \country{Canada}}
\email{arniebh@google.com}
\author{Siyan Zheng}
\authornotemark[1]
\affiliation{%
  \institution{Google}
  \city{Mountain View}
  \state{CA}
  \country{USA}}
\email{heatherzheng@google.com}
\author{Anlan Yu}
\authornotemark[1]
\affiliation{%
  \institution{Google}
  \city{Mountain View}
  \state{CA}
  \country{USA}}
\email{anlanyu@google.com}
\author{Palaksh Rungta}
\affiliation{%
  \institution{Google}
  \city{Mountain View}
  \state{CA}
  \country{USA}}
\email{palaksh@google.com}
\author{Jiawei Li}
\affiliation{%
  \institution{Google}
  \city{Mountain View}
  \state{CA}
  \country{USA}}
\email{jweili@google.com}
\author{Changping Meng}
\affiliation{%
  \institution{Google}
  \city{Mountain View}
  \state{CA}
  \country{USA}
}
\email{changping@google.com}
\author{Dapeng Hong}
\affiliation{%
  \institution{Google}
  \city{Mountain View}
  \state{CA}
  \country{USA}}
\email{dapengh@google.com}
\author{Chuan He}
\affiliation{%
  \institution{Google}
  \city{Mountain View}
  \state{CA}
  \country{USA}}
\email{hec@google.com}
\author{Onkar Dalal}
\affiliation{%
  \institution{Google}
  \city{Mountain View}
  \state{CA}
  \country{USA}}
\email{onkardalal@google.com}

\renewcommand{\shortauthors}{Arnab Bhadury et al.}

 \date{Apr 2026}
\begin{abstract}
Large-scale recommender systems, particularly short-form video platforms, are often bottlenecked by massive popularity feedback loops. In such environments, as models recommend popular items, they generate an overwhelming amount of skewed training data for "head" items. This creates a self-reinforcing cycle where retrieval and ranking models memorize "head" item patterns at the expense of generalizing across the vast "tail" of the catalogue. While Curriculum Learning (CL) offers a powerful mechanism to break this feedback loop by systematically exposing models to progressively more difficult and less frequent examples, its adoption in industrial recommendation has been hampered by hardware utilization inefficiencies or the needs for complicated pre-processing techniques because dynamic data rejection algorithms tend to starve hardware accelearators (TPUs/GPUs) by becoming largely CPU-bound. In this work, we introduce a scalable Soft Curriculum Learning framework designed specifically for continuous training setups within industry-scale retrieval and ranking models. By utilizing loss annealing and in-graph weight adjustments rather than rigid data filtering, we break the popularity feedback and enable dynamic curriculum pacing without sacrificing system throughput. We demonstrate empirical evidence through applications across sequence-based retrieval models (such as SASRec), two-tower retrieval models, and large-scale continuous ranking models. Online A/B tests on our short-video platform demonstrate substantial lifts in both overall user satisfaction and fresh content consumption, all without degrading model throughput.
\end{abstract}




\copyrightyear{2026}

\acmYear{2026}
\setcopyright{cc}
\setcctype{by}
\acmConference[RecSys '26]{20th ACM Conference on Recommender Systems}{September 27-October 02, 2026}{Minneapolis, MN, USA}
\acmBooktitle{20th ACM Conference on Recommender Systems (RecSys '26), September 27-October 02, 2026, Minneapolis, MN, USA}
\acmDOI{10.1145/3773078.3831923}
\acmISBN{979-8-4007-2284-4/2026/09}

\maketitle

\section{Introduction}

Modern recommendation ecosystems are tasked with serving billions of queries while maintaining freshness, fairness, and a high degree of exploration. However, achieving robust generalization remains fundamentally difficult. A pervasive and critical issue in these systems is the popularity bias \cite{chang2024cluster}, which quickly degrades into a massive, self-reinforcing feedback loop. As the system surfaces globally popular content, users naturally engage with it, generating a disproportionate amount of positive training data for these "head" items. The model then learns to over-index on these dense, easily recognizable patterns, further amplifying their recommendation frequency. This feedback loop forces the model into a state of memorization, suppressing fresh uploads, cold-start items, and niche "tail" content.

To foster ecosystem growth and ensure fairness for emerging creators, we require a strong regularization mechanism that forces the model to step outside this feedback loop and learn discriminative patterns for lower-viewed, harder-to-predict items. Curriculum Learning (CL) \cite{bengio2009curriculum} offers a biologically inspired optimization strategy to address this directly. By initially presenting the model with "easy" examples (e.g., highly popular items with dense signals) to find a good initialization state, and gradually shifting the optimization focus to "hard" examples (e.g., tail content or sparse-data users), CL effectively counters popularity bias and prevents item embeddings from collapsing into locally popular optima.

Despite its proven efficacy in domains like Computer Vision \cite{hacohen2019power}  \cite{srinivasan2023curriculum}, Robotics \cite{zhang2020automatic} and Natural Language Processing \cite{cirik2016visualizing}, Curriculum Learning remains notably under-explored within large-scale industrial Information Retrieval (IR) and recommender systems. Most prior research has been confined to smaller-scale datasets or offline academic preference alignment \cite{lai2024adaptive} scenarios, largely due to the severe infrastructure challenges posed by deploying true Curriculum Learning via dynamic data rejection in continuously trained, massive-scale pipelines. In this paper, we propose a scalable alternative: \textbf{Soft Curriculum Learning}. Instead of discarding training examples, we leverage cyclic and annealed loss weighting directly on the TPU core to modulate difficulty and break the feedback loop. Specifically, we define the difficulty of an item and cyclically increase the weight of "difficult" items. We also developed two curriculum weight strategies and evaluate the effectiveness on sequence-based retrieval models, two-tower retrieval models and large-scale continuous-trained ranking models. Both online A/B test results and offline results show that our methods help boost engagement and fresh content consumption. 

Consequently, this work serves as a comprehensive case study demonstrating how CL techniques can be successfully adapted and scaled for production-grade recommendation environments, providing a vital blueprint for future industry deployments.
\section{Related Work}

The foundational concept of Curriculum Learning (CL) was formalized by \citet{bengio2009curriculum}. Drawing parallels to human education, they demonstrated that training models on a sequence of progressively harder examples yields faster convergence and guides the optimizer toward better local minima, particularly in non-convex loss landscapes. Extensive literature in the RecSys community has documented the dangers of feedback loops \cite{sinha2016deconvolving}. Works exploring counterfactual risk minimization \cite{swaminathan2015counterfactual} and popularity de-biasing \cite{chang2024cluster} highlight that models trained on logged bandit feedback inevitably amplify existing biases unless explicitly corrected. 

Curriculum Learning intersects with these efforts by providing a dynamic, schedule-based approach to de-biasing, rather than static propensity weighting. The application of CL to Information Retrieval and Recommender Systems has gained traction in recent years, though it remains an under-explored area compared to other deep learning fields.
\citet{penha2020challenges} conducted an empirical study on CL strategies for conversation response ranking, demonstrating that sorting training data by difficulty (estimated via a supervised scoring function) directly improves neural ranking retrieval effectiveness. 
\citet{macavaney2020efficient} applied training curricula for open-domain answer re-ranking. Their work proposed several heuristics to estimate sample difficulty, down-weighting difficult samples early in the training process, which yielded significant improvements in Mean Reciprocal Rank (MRR) for models like BERT and ConvKNRM.
\citet{wang2024curriculum} introduced a Unified Meta-Learning Framework for Fair Ranking, demonstrating that curriculum learning schedulers can successfully guide a meta-learner to gradually mitigate data skewness and address historical exposure bias in ranking algorithms. 

Our work builds upon these theoretical and offline ranking foundations but adapts them to bypass the computational bottlenecks of continuous, large-scale industrial training.

\section{Difficulties of Curriculum Learning in Practice}

While theoretical frameworks for Curriculum Learning are elegant and well-suited for breaking feedback loops, translating them into massive, continuously trained recommendation systems introduces distinct engineering and optimization hurdles.

\subsection{CPU-Bound Bottleneck}

An ideal, "true" curriculum learner relies on dynamic data rejection. In this paradigm, a filter is applied to the training dataset on the fly to build mini-batches that adhere to a specific difficulty threshold at a given training step. However, modern deep learning architectures execute data pre-processing (such as parsing and filtering) primarily on CPUs before feeding the batches into high-performance Tensor Processing Units (TPUs) or Graphic Processing Units (GPUs).

If a strict curriculum policy dictates that a large volume of "easy" or "popular" candidates must be rejected to focus on the tail, the queue buffer inevitably empties faster than the CPU can refill it. As a result, the hardware accelerator is starved of data, plummeting the system's overall throughput. Implementing dynamic data rejection without starvation would require an entirely separate pipeline acting as a massive, parallel mini-batch builder, which can be prohibitively difficult in large industry scale recommendation systems.

\subsection{The Pitfalls of Global Curriculum Pacing}
\emph{Global curriculum learning} strategy forces models to exclusively train on easy samples at the beginning and hard ones at the end of the global life cycle.
While this approach has demonstrated success in academic literature, those successes are largely confined to one-time training paradigms. In contrast, large-scale industrial recommendation systems rely on continuous training, where the model is constantly updated. Through time, the training data distribution shifts from day to day, and a global curriculum that permanently transitions to focusing only on "hard" examples fails to capture the current state of the ecosystem.
Constantly removing "easy" or popular samples after a specific turning point prevents the model from adapting to newly emerging popular trends. We discuss such limit of global curriculum pacing with empirical results and performance comparison in \cref{subsec:global-cv-results}.


\section{Methodology: Soft Curriculum Learning}

To circumvent CPU bottlenecks, mitigate catastrophic forgetting, and effectively combat popularity bias, we designed a Soft Curriculum Learner.

\subsection{Loss Annealing}

Rather than filtering the dataset, the soft curriculum approach passes all data through the pipeline without rejection, keeping the pipeline entirely bound by TPU/GPU compute. To enforce the curriculum, we apply dynamic loss annealing. The loss vector is multiplied by an annealed weight coefficient that adjusts based on training progress and item difficulty. Because this scalar multiplication executes directly on the TPU/GPU core, it introduces virtually no overhead.

\subsection{Quantifying Task Difficulty}\label{sec:diff}

To break the popularity feedback loop, we must accurately identify which items are benefiting from it. We formalize difficulty $d$ using two primary metrics: 

\begin{itemize}
    \item Engagement-based: Targeting on resolving popularity bias, we directly borrow the raw engagement signals, and assign lower difficulty to items with higher engagement volumes.
    
    \item Uncertainty-based (SNGP Variance): To move beyond pure popularity heuristics, we quantify the prediction uncertainty using Spectral-normalized Neural Gaussian Process (SNGP)\cite{Liu:2020}. A large SNGP uncertainty indicates the model is less confident about the prediction due to lack of information. This is naturally compatible to the definition of difficulty level to CL. We define a normalized SNGP uncertainty-inferred difficulty.
    
\end{itemize}

\subsection{Cyclic Curriculum Pacing}

Instead of defining a global curriculum that progresses continuously, we implemented a Cyclic Curriculum that resets periodically. 

To align with our continuous training pipeline, we partition the data stream into discrete sub-day intervals, where the model trains on a fixed volume of $X$ examples during each interval.

We define training progress $p \in [0, 1]$ within each sub-day interval as
$p = \text{step} \bmod X/X$, where $\text{step}$ represents the cumulative number of training examples processed.


\subsection{Curriculum Weight Strategies}

We investigated two distinct curriculum learning strategies. In both approaches, $p \in [0,1]$ represents the training progress , $d \in [0,1]$ represents the estimated difficulty of the example, and $\alpha$ serves as a hyperparameter to tune the percentage of hard and easy examples. The curriculum weight $w$ for a given sample is computed using one of the following methods:

\paragraph{1. Hard-Example Warmup}
Our first strategy isolates hard examples, systematically down-weighting their influence during the early stages of optimization while maintaining easy examples at a constant baseline weight. The weight is calculated as:
\begin{equation}
    w = 1.0 - \alpha\cdot d\cdot (1.0 - p).
    \label{eq:hard_warmup}
\end{equation}
Under this formulation, at the onset of training ($p=0$), the hardest examples ($d=1$) are assigned a penalized initial weight of $1.0 - \alpha$. As training advances ($p \rightarrow 1.0$), the penalty term $\alpha \cdot d (1.0 - p)$ monotonically decays to zero. Thus, this mechanism safely and smoothly introduces harder examples over time until the penalty vanishes entirely, ensuring that the more-mature network trains over harder examples with $w=1.0$.

\paragraph{2. Bidirectional Difficulty Scaling}
Our second strategy goes a step further by dynamically shifting the network's focus from easy to hard examples over the course of training. We calculate this weight using:
\begin{equation}
    w = 1.0 + \alpha\cdot (2p - 1)\cdot(2d - 1).
    \label{eq:bidirectional_scaling}
\end{equation}
This creates a linear inversion of priority. At the start of training ($p=0$), easy examples ($d=0$) receive an amplified weight of $1.0 + \alpha$, while hard examples ($d=1$) are suppressed to $1.0 - \alpha$. By the end of training ($p=1$), this distribution perfectly inverts: easy examples decay to $1.0 - \alpha$, and hard examples are emphasized at $1.0 + \alpha$. 

\section{Experimental Results}

\subsection{Ranking Evaluation}\label{sec:ranking_evals}

\subsubsection{Offline Evaluation}
In our baseline, examples are weighted inversely to their popularity, assigning higher weights to items with fewer engagements. To implement curriculum learning, we replaced this static popularity factor with a dynamic curriculum weight. To ensure the observed improvements were driven by the curriculum strategy and not merely an artifact of discarding the popularity penalty, we evaluated an ablation model with the popularity factor completely removed; this yielded neutral offline results and negative online results, confirming the efficacy of our proposed weighting. Because standard Area Under the ROC Curve (AUC) is incomparable when training with dynamic sample weights, we evaluated offline performance using an unweighted AUC metric. 

In \cref{tab:experiment_results}, we compare the following variants with the production models.

\begin{itemize}
    \item Engagement-Hard (Eng-Hard): Difficulty calibrated by historical engagement-count distributions and use hard example warmup scaling.
    \item SNGP-Hard: Difficulty calibrated via Spectral-normalized Neural Gaussian Process uncertainty and use hard example warmup scaling.
    \item Eng-Bi: Difficulty calibrated by historical engage-count distributions and use bidirectional difficulty scaling.
    \item SNGP-Bi: Difficulty calibrated via Spectral-normalized Neural Gaussian Process uncertainty and use bidirectional difficulty scaling.
    
\end{itemize}

We define difficulty $d$ as
\begin{equation}
    d = 
    \begin{cases}
        \exp(-quantile(\text{engagement})), & \text{For engagement-based}, \\
        quantile(\hat{U}_{\text{SNGP}}), & \text{For SNGP-based.}
    \end{cases}
\end{equation}

The result shows that both SNGP and Engagement-based curriculum methods surpasses baseline in all tasks. Employing a bidirectional difficulty curriculum achieves superior overall performance compared to a standard hard-example warm-up approach. AUC improvements of treatment Eng-bi across varying video ages is presented in \cref{tab:freshness_analysis}. The data indicates that "cold" and new content has the highest marginal gain by curriculum learning.



\begin{table}
\centering
\caption{Ranking Offline Performance Evaluations}
\label{tab:experiment_results}
\begin{tabular}{lcccc}
\toprule
Task &  Eng-Hard & Eng-Bi & SNGP-Hard & SNGP-Bi \\
\midrule
Imp. Pos. & +0.02\% & \textbf{+0.03\%} & +0.01\% & +0.01\% \\
Exp. Neg. & +0.06\% & \textbf{+0.08\%} & +0.02\% & +0.06\% \\
Satisfaction & +0.02\% & \textbf{+0.04\%} &+0.01\%& +0.02\%\\
Imp. Neg. &  +0.03\% & \textbf{+0.04\%} & +0.01\% &+0.02\% \\
\bottomrule
\end{tabular}
\end{table}

\begin{table}
\centering
\begin{minipage}{\columnwidth}
\caption{Performance Comparison Categorized by Video Age of Eng-Bi}
\label{tab:freshness_analysis}
\begin{tabular}{lcccccc}
\toprule
Video Age\protect\footnote{Number of Video Age (in days) from Bucket 1 to 5 in ascending order.} & Imp. Pos. & Exp. Neg. & Satisfaction & Imp. Neg \\
\midrule
Age Bucket 1 & \textbf{+0.05\%} & \textbf{+0.14\%} & \textbf{+0.05\%} & \textbf{+0.08\%} \\
Age Bucket 2 & +0.02\% & +0.08\% & +0.02\% & +0.03\% \\
Age Bucket 3 & +0.02\% & +0.05\% & +0.02\%  & +0.02\% \\
Age Bucket 4 & +0.02\% & +0.06\% & +0.02\%  & +0.03\% \\
Age Bucket 5 & +0.02\% & +0.05\% & +0.02\% & +0.02\% \\
\bottomrule
\end{tabular}
\end{minipage}
\end{table}
\subsubsection{Online A/B Testing}
We deployed the Soft CL Ranker in a large-scale A/B test on a short-video recommendation surface for 7 days. 
Table \ref{tab:online_metrics_user} and table \ref{tab:online_metrics_views} compared the production baseline against two variants of our Soft CL Ranker: bidirectional engagement based Soft CL, bidirectional SNGP based Soft CL. Compared to the production baseline, both Soft CL variants demonstrated statistically significant improvements across primary engagement metrics, confirming the robustness of the soft curriculum approach regardless of the underlying difficulty heuristic.  
Notably, the Bidirectional engagement-based Soft CL variant had the most substantial gains. This variant achieved a +0.54\% increase in User Satisfaction and +5.7\% increase in cold-start user engagement, in the freshest age bucket.

\begin{table}
\centering
\caption{Online Ranking Model Performance: User Metrics}
\label{tab:online_metrics_user}
\small
\begin{tabular}{lccc}
\toprule
Method & User Satisfaction & User Growth & User Retention \\
\midrule
SNGP-bi & +0.42\% & \textbf{+0.15\%} & \textbf{+0.15\%} \\
Eng-bi & \textbf{+0.54\%} & +0.11\% & \textbf{+0.15\%} \\
\bottomrule
\end{tabular}
\end{table}

\begin{table}
\centering
\caption{Online Ranking Model Performance: Engagements and Cold-Start Engagement Metrics}
\label{tab:online_metrics_views}
\small
\begin{tabular}{lcccc}
\toprule
\multirow{2}{*}{Method} & User & \multicolumn{3}{c}{Cold-start Engagement} \\
\cmidrule(lr){3-5}
 & Engagement & Age Bucket 1 & Age Bucket 2 & Age Bucket 3 \\
\midrule
SNGP-bi & +0.92\% & +4.71\% & +1.13\% & +1.54\% \\
Eng-bi & \textbf{+1.89\%} & \textbf{+5.70\%} & \textbf{+2.99\%} & \textbf{+2.72\%} \\
\bottomrule
\end{tabular}
\end{table}


\subsubsection{Comparison with Global Curriculum Learning}
\label{subsec:global-cv-results}
Global Curriculum Learning approach adjusts difficulty incrementally on a day-by-day basis. We evaluated a global curriculum learning strategy by applying data rejection at specific progression milestones. The following configurations were evaluated:
\begin{itemize}
    \item \textbf{Baseline:} The production model without data filtering.
    \item \textbf{Global-1Day:} Filtering out easy examples after 1 day of training.
    \item \textbf{Global-3Day:} Filtering out easy examples after 3 days of training.
\end{itemize}

Despite the lower CPU cost for the daily-base sampling rejection by filtering materialzation, both the offline and online results for the global curriculum approaches were distinctly negative, as summarized in Table \ref{tab:global_offline_cl_results} and Table \ref{tab:global_online_cl_results}. By permanently removing easy examples from the training distribution at these day-level intervals, model degrades overall generalization.

 \begin{table}
\centering
\caption{Global Curriculum Learning Offline Performance Evaluations}
\label{tab:global_offline_cl_results}
\begin{tabular}{lccccc}
\toprule
Task & Imp. Pos. & Exp. Neg. & Satisfaction & Imp. Neg.\\
\midrule
Global-1Day & -0.15\% & -0.30\% & 0.0\%  & -0.19\% \\
Global-3Day & -0.19\% & -0.09\% & -0.03\% & -0.27\%\\
\bottomrule
\end{tabular}
\end{table}

\begin{table}
\centering
\caption{Global Curriculum Learning Online Performance Evaluations}
\label{tab:global_online_cl_results}
\small
\begin{tabular}{lccc}
\toprule
Task & User Satisfaction & User Growth & User Retention \\
\midrule
Global-1Day & -0.71\% & -0.19\% & -0.12\% \\
Global-3Day & -0.72\% & -0.22\% & -0.17\% \\
\bottomrule
\end{tabular}
\end{table}

\subsection{Retrieval Evaluation}
To comprehensively evaluate the impact of Soft Curriculum Learning across the large-scale industrial recommender system stack, we conduct experiments on two types of retrieval models inspired from: 1) SASRec-like Sequence-based Architecture, and 2) Two-Tower Architecture \cite{yi2019sampling}.

Note that SNGP uncertainty is calculated at serving time within the ranking model. Calculating SNGP for the global item population during retrieval-stage training is computationally expensive. Therefore, we evaluate only engagement-based metrics in the retrieval stage.

\subsubsection{SASRec Style Model} We evaluate our proposed curriculum weighting strategy on the sequence-based retrieval model by employing the engagement-based difficulty metric and the Hard-Example Warmup strategy formulated in \cref{eq:hard_warmup}. 
During experimentation, we vary the hyperparameters $\alpha$, $frequency$, and select the combination that achieves the best performance.

\paragraph{Offline Evaluation}
We use mean average precision (MAP@K) to measure the offline performance for the SASRec-style retrieval model. MAP is calculated as a mean across the average precision of user-video predictions over the evaluation batch size $B$. For each user query $i \in \{1, \dots, B\}$, let $\text{rank}_{\mathcal{V}}(i)$ denote the rank of the true positive video among a set of videos sampled from the global population $V$. The evaluation MAP@K is computed as:
\begin{equation}
\label{eq:map_sasrec}
\text{MAP@}K_{\text{SASRec}} = \frac{1}{B} \sum_{i=1}^B \mathbb{I}\big(\text{rank}_{\mathcal{V}}(i) \le K\big) \cdot \frac{1}{\text{rank}_{\mathcal{V}}(i)}
\end{equation}
where $\mathbb{I}(\cdot)$ is the indicator function.

For all variations evaluated in this experiment, the offline MAP shows a marginal decline compared to the baseline, as reported in \cref{tab:sasrec_offline}. This discrepancy stems from the fundamental difference in model capacity and feature density between retrieval and ranking stages. Our SASRec-style retrieval model, designed for extreme efficiency in massive-scale candidate generation, operates on a much leaner feature set than its ranking counterpart.

By devoting this limited model capacity to properly exploring long-tail items through our curriculum weighting strategy, the treatment model inevitably trade-offs some representation accuracy on the popular head. Crucially, because the offline test set is highly skewed and contains sparse tail interactions, our model's improved capability on the tail is barely reflected in the overall MAP. Conversely, its slight drop in head-item accuracy is heavily penalized by the head-heavy distribution. Consequently, the baseline model achieves a deceptively higher MAP simply by under-exploring the tail and overfitting to the dominant head items, a short-term precision gain that does not necessarily translate to a healthier retrieval diversity.

\bigskip
\begin{table}
\centering
\caption{SASRec Retrieval Offline Performance Evaluations}
\label{tab:sasrec_offline}
\begin{tabular}{lcc}
\toprule
Task & Baseline & Eng-Hard \\
\midrule
$\text{MAP@1}_{\text{SASRec}}$  & \textbf{0.009537} & 0.009123 \\
$\text{MAP@5}_{\text{SASRec}}$  & \textbf{0.01753}  & 0.01684  \\
$\text{MAP@10}_{\text{SASRec}}$ & \textbf{0.02012}  & 0.01939  \\
$\text{MAP@20}_{\text{SASRec}}$ & \textbf{0.02202}  & 0.02128  \\
$\text{MAP@50}_{\text{SASRec}}$ & \textbf{0.02364}  & 0.02290  \\
\bottomrule
\end{tabular}
\end{table}

\paragraph{Online A/B Testing}
As demonstrated in \cref{tab:retrieval_online} over a period of two weeks of experimentation, our Eng-Hard Soft CL variant significantly outperforms the production baseline. Even though our offline metrics are less comparable to the baseline, we do not observe topline engagement loss. Notably, unique engagements from this retrieval model increased by $7.7\%$, confirming the enhanced quality of the retrieval model. Furthermore, the sustained gains in engagements across different age buckets indicate that the Soft CL framework effectively promotes and accelerates the discovery of cold-start content. 

\begin{table}
\centering
\caption{SASRec Retrieval Online A/B Performance Evaluations}
\label{tab:retrieval_online}

\begin{tabular}{l|c|ccc}
\toprule
\multirow{2}{*}{Method} & Retrieval & \multicolumn{3}{c}{Cold-start engagement} \\
\cline{3-5}
 & Engagements & \shortstack{Age\\Bucket 1} & \shortstack{Age\\Bucket 2} & \shortstack{Age\\Bucket 3} \\
\hline
Eng-Hard & +7.7\% & +0.27\% & +0.69\% & +0.61\% \\
\bottomrule
\end{tabular}
\end{table}

\subsubsection{Two-Tower Style Model} 
Similar to the setup of SASRec-style retrieval model, we use Eng-Hard type solution for our two-tower style retrieval model. During experimentation, we vary the hyperparameters $\alpha$, $frequency$, and select the combination that achieves the best performance for the model.

For offline evaluation, we also adopt the Mean Average Precision (MAP) as defined in \cref{eq:map_sasrec}. However, as this model utilizes in-batch negative sampling during both training and evaluation, the candidate video pool is constrained to the evaluation batch size $B$ rather than drawn from the global video population like our SASRec-style retrieval model does. This discrepancy in the candidate pool size and composition leads to a shift in metric scales between the two models. Consequently, we denote this specific offline metric as $\text{MAP@}K_{\text{TwoTower}}$.

\cref{tab:two_tower_offline} presents the offline comparison between the baseline and our Eng-Hard based Soft CL method. Notably, the treatment model achieves a higher $\text{MAP@}K_{\text{TwoTower}}$ than the baseline, contrasting with the trend observed in the SASRec-style retrieval model. We attribute this to the feature-rich, generalized architecture of the Two-Tower model. By gradually introducing harder examples as training progressed, the network was able to learn richer, more robust feature representations within the shared video embedding space. This hypothesis is consistent with our findings in \cref{sec:ranking_evals}; since ranking models possess even greater feature capacity than retrieval models, they demonstrate a superior ability to generalize when exposed to progressive curriculum difficulty.

\bigskip
\begin{table}
\centering
\caption{Two-Tower Retrieval Offline Performance Evaluations}
\label{tab:two_tower_offline}
\begin{tabular}{lcc}
\toprule
Task & Baseline & Eng-Hard \\
\midrule
$\text{MAP@1}_{\text{TwoTower}}$  & 0.16348 & \textbf{0.17133} \\
$\text{MAP@5}_{\text{TwoTower}}$  & 0.29806  & \textbf{0.32609}  \\
$\text{MAP@50}_{\text{TwoTower}}$ & 0.33235  & \textbf{0.36547}  \\
$\text{MAP@200}_{\text{TwoTower}}$ & 0.33316  & \textbf{0.36566}  \\
\bottomrule
\end{tabular}
\end{table}

\subsection{Production Challenges: Mitigating A/A Variance via Batch Difficulty Normalization}

 A/A variance refers to the statistical fluctuations in model metrics when evaluating identical model configurations. Maintaining low A/A variance is critical, as high variance can obscure genuine algorithmic improvements in the future model iterations.

While Soft Curriculum Learning framework effectively breaks the popularity feedback loop, it natively introduces higher A/A variance. This instability occurs because the curriculum weights are dynamically calculated based on the inherent difficulty $d$ of the items within each randomly sampled batch. Depending on the random distribution of the data, one batch might contain a disproportionate number of "hard" tail items, while the next might be predominantly composed of "easy" head items. Because our weighting strategies scale the loss based on these difficulty scores, this batch imbalance causes the aggregate gradient magnitude to fluctuate wildly from step to step. This fluctuating effective learning rate injects noise into the optimization process, directly inflating A/A variance.

To stabilize the training process and reduce this variance, we normalize the batch level difficulty. For each mini-batch of size $N$, rather than using the raw weight scores $w$, we normalize the curriculum weight of each example such that the mean weight of the batch, $\bar{w}$, is exactly $1$. We ensure that the average curriculum weight applied to any given batch remains strictly consistent, regardless of the random sampling distribution of that specific batch. This normalization effectively stabilizes the gradient updates and significantly mitigates A/A variance.

\section{Conclusion}
We presented an industrial implementation of Soft Curriculum Learning tailored for continuously trained retrieval and ranking models. By employing cyclic schedules and loss annealing, we bypass the CPU bottlenecks associated with traditional Curriculum Learning. Our methodology shows that dynamically adjusting item difficulty using popularity quantiles or SNGP variance acts as a highly effective regularizer in both the retrieval and ranking stages.

In our SASRec-style model, devoting model capacity to learning long-tail, harder examples successfully promotes exploration, yielding a +7.7\% increase in unique retrieval engagements during online A/B testing despite lower offline MAP on head-heavy test sets. In feature-rich, generalized architectures (Two-Tower retrieval and deep ranking models), gradually introducing harder examples helps the network learn richer, more robust representations across the shared embedding space. For ranking models, which possess even greater feature capacity, this generalization translates into broad offline AUC improvements across all multi-task prediction heads. Online A/B tests confirm these gains, demonstrating a +0.54\% increase in overall User Satisfaction and a +5.70\% surge in cold-start content engagement. The result of applying Soft CL is a more well-calibrated system that effectively balances exploration, significantly boosting the distribution of cold-start content without sacrificing overall ecosystem engagement or hardware efficiency.

\bibliographystyle{ACM-Reference-Format} 
\bibliography{library} 
\end{document}